\pdfoutput=1

\documentclass[11pt,twoside,a4paper,cmspaper,final,collab]{cms-tdr}
\def\svnVersion{166732}\def\cmsCernNo{2012-089}\def\cmsCernDate{\today}\def\cmsMessage{Submitted to Physics Letters B}

\begin{document}\cmsNoteHeader{HIN-11-010}

\hyphenation{had-ron-i-za-tion}
\hyphenation{cal-or-i-me-ter}
\hyphenation{de-vices}

\RCS$Revision: 165940 $
\RCS$HeadURL: svn+ssh://svn.cern.ch/reps/tdr2/papers/HIN-11-010/trunk/HIN-11-010.tex $
\RCS$Id: HIN-11-010.tex 165940 2013-01-21 15:11:24Z alverson $
\newlength\cmsFigWidth
\ifthenelse{\boolean{cms@external}}{\setlength\cmsFigWidth{0.85\columnwidth}}{\setlength\cmsFigWidth{0.4\textwidth}}
\ifthenelse{\boolean{cms@external}}{\providecommand{\cmsLeft}{top}}{\providecommand{\cmsLeft}{left}}
\ifthenelse{\boolean{cms@external}}{\providecommand{\cmsRight}{bottom}}{\providecommand{\cmsRight}{right}}
\providecommand{\mubinv}{\ensuremath{\,\mu\text{b}^{-1}}\xspace}
\newcommand{\mbinv}{\ensuremath{\,\text{mb}^{-1}}\xspace}
\providecommand {\rd}       {\ensuremath{\mathrm{d}}}
\newcommand {\re}       {\ensuremath{\mathrm{e}}}
\newcommand {\roots}    {\ensuremath{\sqrt{s}}}
\newcommand {\rootsNN}  {\ensuremath{\sqrt{s_{_{NN}}}}}
\newcommand {\dndy}     {\ensuremath{\d N/\d y}}
\newcommand {\dnchdy}   {\ensuremath{\d N_{\mathrm{ch}}/\d y}}
\newcommand {\dndeta}   {\ensuremath{\d N/\d\eta}}
\newcommand {\dnchdeta} {\ensuremath{\d N_{\mathrm{ch}}/\d\eta}}
\newcommand {\dndpt}    {\ensuremath{\d N/\d\pt}}
\newcommand {\dnchdpt}  {\ensuremath{\d N_{\mathrm{ch}}/\d\pt}}
\newcommand {\deta}     {\ensuremath{\Delta\eta}}
\newcommand {\dphi}     {\ensuremath{\Delta\phi}}
\newcommand {\AJ}    {\ensuremath{A_{J}}}
\newcommand {\pp}    {\mbox{pp}\xspace}
\providecommand {\ppbar} {\mbox{p\={p}}}
\providecommand {\pbarp} {\mbox{p\={p}}}
\newcommand {\PbPb}  {\ensuremath{\text{PbPb}}\xspace}
\newcommand {\ak}    {anti-$k_\text{T}$}

\newcommand {\ptg}    {\ensuremath{p_{\mathrm{T}}^\cPgg}\xspace}
\newcommand {\ptj}    {\ensuremath{p_{\mathrm{T}}^\text{Jet}}\xspace}
\newcommand {\phig}    {\ensuremath{\phi^\cPgg}\xspace}
\newcommand {\phij}    {\ensuremath{\phi^\mathrm{Jet}}\xspace}

\newcommand {\npart}  {\ensuremath{N_{\text{part}}}\xspace}

\newcommand{\m}{\ensuremath{\,\text{m}}\xspace}

\newcommand {\naive}    {na\"{\i}ve}
\providecommand{\GEANT} {{Geant}\xspace}
\providecommand{\PHOJET} {\textsc{phojet}\xspace}

\def\d{\mathrm{d}}

\providecommand{\PKzS}{\ensuremath{\mathrm{K^0_S}}}
\providecommand{\Pp}{\ensuremath{\mathrm{p}}}
\providecommand{\Pap}{\ensuremath{\mathrm{\overline{p}}}}
\providecommand{\PgL}{\ensuremath{\Lambda}}
\providecommand{\PagL}{\ensuremath{\overline{\Lambda}}}
\providecommand{\PgS}{\ensuremath{\Sigma}}
\providecommand{\PgSm}{\ensuremath{\Sigma^-}}
\providecommand{\PgSp}{\ensuremath{\Sigma^+}}
\providecommand{\PagSm}{\ensuremath{\overline{\Sigma}^-}}
\providecommand{\PagSp}{\ensuremath{\overline{\Sigma}^+}}

\newcommand {\photonjet}{photon+jet}
\let\gammajet\photonjet

\providecommand{\ptjet}{\ptj}
\providecommand{\ptgamma}{\ptg}
\providecommand{\ptphoton}{\ptg}
\providecommand{\rjg}{\ensuremath{R_{J\cPgg}}\xspace}
\providecommand{\xjg}{\ensuremath{x_{J\cPgg}}\xspace}
\providecommand{\sjg}{\ensuremath{\sigma(\Delta\phi_{J\cPgg})}\xspace}
\providecommand{\ptg}{\ensuremath{p_{\mathrm{T},\cPgg}}\xspace}
\providecommand{\dphijg}{\ensuremath{\Delta\phi_{J\cPgg}}\xspace}
\providecommand{\avexjg}{\ensuremath{\langle x_{J\cPgg}\rangle}\xspace}
\providecommand{\sigeta}{\ensuremath{\sigma_{\eta\eta}}\xspace}
\providecommand{\HYDJET} {\textsc{hydjet}\xspace}
\providecommand{\PYTHIAHYDJET} {\textsc{pythia + hydjet}\xspace}
\providecommand{\PYQUEN} {\textsc{pyquen}\xspace}
\cmsNoteHeader{HIN-11-010} % This is over-written in the CMS environment: useful as preprint no. for export versions

\title{Studies of jet quenching using isolated-photon+jet correlations in PbPb and pp collisions at $\sqrt{s_{NN}} = 2.76\TeV$}

\date{\today}

\abstract{
Results from the first study of isolated-photon+jet correlations in
relativistic heavy ion collisions are reported. The analysis uses data
from \PbPb\ collisions at a centre-of-mass energy of 2.76\TeV per nucleon pair
corresponding to an integrated luminosity of 150\mubinv
recorded by the CMS experiment at the LHC. For events containing an isolated
photon with transverse momentum $\ptg > 60\GeVc$ and an associated jet with $\ptj >
30\GeVc$, the \photonjet\ $\pt$ imbalance is studied as a function of
collision centrality and compared to \pp\ data and \PYTHIA
calculations at the same collision energy.
Using the $\ptg$ of the isolated photon as an estimate of the momentum of the associated parton at production,
this measurement allows a characterisation of the in-medium parton energy loss.
For more central \PbPb\ collisions, a significant decrease in the ratio  $\ptj/\ptg$ relative
to that in the \textsc{pythia} reference is observed. Furthermore, significantly
more $\ptg > 60\GeVc$ photons in \PbPb\ are observed not to have an
associated $\ptj >
30\GeVc$ jet, compared to the reference.
However, no significant broadening of the \photonjet\
azimuthal correlation is observed.
}

\hypersetup{%
pdfauthor={CMS Collaboration},%
pdftitle={Studies of jet quenching using isolated-photon+jet correlations in PbPb and pp collisions at sqrt(s[NN]) = 2.76 TeV},%
pdfsubject={CMS},%
pdfkeywords={CMS, physics, heavy ion, photon}}

\maketitle %maketitle comes after all the front information has been supplied

\section{Introduction}

Parton scatterings with large momentum transfer produce energetic
particles which can be used as ``probes'' to study the strongly
interacting medium
created in high-energy heavy ion collisions~\cite{Shuryak:1978ij,Shuryak:2004cy}. The
production of high transverse momentum (\pt{}) partons and photons in ``hard'' processes occurs
over very short time scales, $\tau\approx 1/\pt \lesssim
0.1\unit{fm}/c$, and thus their yields can be potentially modified
by final-state interactions occurring while they traverse the medium. Since the production cross sections
of these energetic particles are calculable
using perturbative quantum chromodynamics, they have long been
recognised as particularly useful ``tomographic'' probes of the created medium~\cite{Appel:1985dq,Blaizot:1986ma,Gyulassy:1990ye,Wang:1991xy,Baier:1996sk,Zakharov:1997uu,Neufeld:2010fj}.

In PbPb collisions at the Large Hadron Collider (LHC), the
effects of the produced medium have been studied using back-to-back
dijets which were observed to be significantly unbalanced in their transverse momenta~\cite{Chatrchyan:2011sx,Aad:2010bu,CMS:2012ni}. The advantage of the large yield of dijets (as compared to \photonjet\ pairs)
is, however, offset by a loss of information about the initial properties of the probes, i.e.\
prior to their interactions with the medium.
Correlating two probes that both undergo energy loss also
induces a selection bias towards scatterings occurring at,
and oriented tangential to, the surface of the medium.
At leading order (LO), photons are produced back-to-back with an associated parton (jet) having close
to the same transverse momentum.
Furthermore, these photons do not strongly interact with the medium.
The yields of isolated photons in \PbPb{} collisions were found to match
the expectation based on \pp{} data and the number of nucleon-nucleon
collisions, with a modification factor of $R_{AA} = 0.99\pm
0.31\text{(stat.)}\pm 0.26\text{(syst.)}$~\cite{HIPhoton}.
Therefore, \photonjet\ production has been hailed as the ``golden channel'' to
investigate energy loss of partons in the medium \cite{Wang:1996yh,Wang:1996pe}.

``Prompt photons'' are photons produced directly in the hard
sub-processes. Experimentally, events with enriched production of prompt photons are
selected using an isolation requirement, namely that the additional energy
in a cone of fixed radius around the direction of the reconstructed
photon be less than a specified value~\cite{HIPhoton}. This restriction
yields ``isolated photons'' ($\gamma$), which consist
mostly of prompt photons produced directly in the initial hard scattering.
Background photons from the decays of neutral mesons, such as $\Pgpz$,
$\Pgh$, and $\omega$, are suppressed by this isolation requirement, as
they are predominantly produced via jet fragmentation.

This Letter describes the first study of the jet energy loss
using isolated-photon+jet pairs from PbPb data at a
nucleon-nucleon centre-of-mass energy $\sqrt{s_{NN}} =
2.76\TeV$. An integrated \PbPb\ luminosity of
$\int\mathcal{L} dt = 150\mubinv$ was collected by the
Compact Muon Solenoid (CMS) experiment during the 2011 running of the LHC.
For comparison, a \pp\ reference dataset with $\int\mathcal{L} dt \approx
200\nbinv$ at $\sqrt{s} = 2.76\TeV$ was obtained
in 2011.

The goal of this analysis is to characterise possible modifications of jet properties as a function of centrality using isolated-\photonjet\ events in \PbPb\ collisions.
The properties of isolated-\photonjet{} pairs are studied
via the azimuthal angular correlation in $\dphijg = |\phij - \phig|$ and the
transverse momentum ratio given by
$\xjg = \ptj/\ptg$.
Photons with transverse momentum of
$\ptg > 60\GeVc$ are selected in a pseudorapidity range of
$\lvert\eta^\gamma\rvert<1.44$, using isolation criteria detailed in Sections \ref{sec:MC} and \ref{sec:photon_reco}.
These photons are then correlated with jets having $\ptj > 30\GeVc$
and $\lvert\eta^\text{Jet}\rvert<1.6$.
Parton energy loss due to induced gluon radiation can lead to a shift of the \xjg\
distribution towards lower values.  In addition, parton energy loss
can cause reconstructed jets to fall below the $\ptj > 30\GeVc$
threshold, leading to a reduction of the fraction of photons with an associated jet.

Section 2 of this Letter begins with a description of the experimental setup as well as the event triggering, selection
and characterisation. The Monte Carlo simulation, the photon and jet reconstruction, and the analysis procedure are also
described. The results and their systematic uncertainties are presented in Section 3, followed by a summary in Section 4.

\section{The CMS detector}
\label{sec_method}

Particles produced in \pp{} and \PbPb{} collisions are studied using
the CMS detector~\cite{JINST}.
The central tracking system is comprised of silicon pixel
and strip detectors that allow for the reconstruction of charged-particle trajectories
in the pseudorapidity range $\lvert\eta\rvert < 2.5$, where $\eta = -\ln[\tan(\theta/2)]$ and $\theta$ is the polar angle relative to the counterclockwise beam direction. Photons are reconstructed using the energy deposited
in the barrel region of the $\mathrm{PbWO_4}$ crystal electromagnetic
calorimeter (ECAL), which covers a pseudorapidity range of
$\lvert\eta\rvert < 1.479$, and has a finely segmented granularity of $\Delta\eta\times\Delta\phi = 0.0174 \times
0.0174$. The brass/scintillator hadron calorimeter (HCAL) barrel region covers $\lvert\eta\rvert < 1.74$, and has a segmentation of $\Delta\eta\times\Delta\phi = 0.087 \times 0.087$. Endcap regions of the HCAL and ECAL extend the $\lvert\eta\rvert$ coverage out to about 3.
The calorimeters and tracking systems
are located within the $3.8\unit{T}$ magnetic field of the super-conducting solenoid.
In addition to the barrel and endcap detectors, CMS includes hadron forward
(HF) steel/quartz-fibre Cherenkov calorimeters, which cover the forward rapidity
of $2.9<|\eta|<5.2$ and are used to determine the degree of overlap (``centrality'')
of the two colliding Pb nuclei~\cite{Chatrchyan:2011pb}. A set of scintillator tiles, the beam scintillator
counters, is mounted on the inner side of each HF for triggering and
beam-halo rejection for both \pp{} and \PbPb{} collisions.

\subsection{Trigger and event selection}
\label{sec_samples}

Collision events containing high-\pt\ photon candidates are
selected online by the CMS two-level trigger system consisting of the Level-1 (L1)  and High Level Trigger (HLT).
First, events are selected using an inclusive single-photon-candidate L1 trigger with a transverse momentum
threshold of 5\GeVc.
Then, more refined photon candidates
are reconstructed in the HLT using a clustering algorithm (identical to that used for offline analysis)
applied to energy deposits in the ECAL.
Events containing  a reconstructed photon candidate with $\ptg > 40\GeVc$ are
stored for further analysis.
This HLT selection
is fully efficient for events containing a photon with $\ptg > 50\GeVc$ and
the analysis presented here includes all photons with $\ptg > 60\GeVc$.

In order to select a pure sample of inelastic hadronic \PbPb\ collisions for analysis,
further offline selections were applied to the triggered event sample similar to \cite{Aad:2010bu}.
Notably among these include requiring a reconstructed event vertex,
and requiring at least 3 calorimeter towers in the HF on both sides of the
interaction point with at least 3\GeV total deposited energy in each tower.
Beam halo events were vetoed based on the timing of the $+z$ and $-z$ BSC signals.
Additionally, events containing HCAL noise~\cite{Chatrchyan:2009hy} are rejected to
remove possible contamination of the jet sample. Details about this event selection
scheme can be found in \cite{Chatrchyan:2011sx}.
The number of events removed by these criteria are shown
in Table~\ref{evselcuts}. Analysis of the Monte Carlo (MC) reference, described in Section~\ref{sec:MC},
uses identical event selection, except for the calorimeter noise
rejection, which is a purely experimental effect.

The online trigger scheme for the \pp\ data at $2.76\:\TeV$ is the same as that used for the CMS \pp\ prompt photon analysis at 7~\TeV\ \cite{CMSppPhoton}. The \pp\ trigger requires at least one reconstructed electromagnetic cluster with a minimum transverse energy of 15~\GeVc. The offline criterion applied to select \pp\ hadronic collision events is similar to previous CMS \pp\ papers \cite{pphadronspectraCMS}.
Apart from the trigger and hadronic collision selection the \pp\ analysis uses the same event selections as the \PbPb\ analysis~\cite{HIPhoton}.

\begin{table*}[htbp]
\topcaption{The impact of the various event-selection criteria on the $\int\mathcal{L}dt = 150\mbinv$
\PbPb{} data sample. In the third column, the percentages are with respect to the line
above. The
selections are applied in the sequence listed. Recall that $\dphijg = |\phij - \phig|$.}
\begin{center}
\begin{tabular}{l|r|r}
Selection & Events remaining & \% of previous\\
\hline
Collision events with a photon of $\ptg > 40\GeVc$ & 252576 & --\\ % it was 96.86
HCAL cleaning & 252317 & 96.76\\   % = 99.90 * 96.86
Isolated photon candidate $\ptg > 60\GeVc, \lvert\eta\rvert < 1.44$ & 2974 & 1.18\\
Jet candidate $\ptj > 30\GeVc, \lvert\eta\rvert < 1.6$ & 2198 & 73.91\\
$\dphijg > \frac{7}{8}\pi$ & 1535 & 69.84\\
\end{tabular}
\end{center}
\label{evselcuts}
\end{table*}

For the analysis of PbPb events, it is important to determine the
degree of overlap between the two colliding nuclei, termed collision centrality.
Centrality is determined using the sum of transverse energy reconstructed in the HF.
The distribution of this total energy is used to divide the event sample into equal percentiles of the total nucleus-nucleus interaction cross section.
These finer centrality bins are then combined into four groups; one containing the 10\% most central events (i.e.\ those which have the smallest impact parameter of the two colliding Pb nuclei and which produce
the highest HF energy); two encompassing the next most central 10--30\% and 30--50\% of the events; and finally one with the remaining 50--100\% peripheral events.
Centrality can also be characterised using the number of nucleons participating in the interaction, \npart (with $\npart = 2$ for \pp{}).
The corresponding \npart{} values for a given centrality range are determined from a Glauber calculation~\cite{Miller:2007ri}. Detector effects are accounted for using a \GEANTfour{} simulation~\cite{geant4} of events generated with a multi-phase transport model ({\sc{ampt}})~\cite{Lin:2004en}.
A detailed description of the centrality determination procedure can be found in \cite{Chatrchyan:2011sx}.

\subsection{Monte Carlo simulation}
\label{sec:MC}

The production of high-\pt{} photons by LO processes and
parton radiation and fragmentation channels with a high-\pt{} photon in the final state
are simulated with \PYTHIA{}~\cite{Sjostrand:2006za} (version 6.422,
tune Z2). Tune Z2 is identical to the Z1 tune described in
\cite{Field:2010bc}, except that Z2 uses the CTEQ6L PDF while Z1 uses
CTEQ5L, and the cut-off for multiple parton interactions, $p_{\perp
  0}$, at the
nominal energy of $\sqrt{s_0} = 1.8\TeV$ is decreased by $0.1\GeVc$. Modifications to account for the isospin effect of the colliding
nuclei, i.e.\ the correct cross section weighting of pp, pn, and
nn subcollisions~\cite{Lokhtin:2005px}, is used.
Events containing isolated photons are selected using the
generator-level information of the \PYTHIA{} events.
The isolation criterion requires that the total energy within a cone
of radius $\Delta R =
\sqrt{(\Delta\eta)^2 + (\Delta\phi)^2} = 0.4$ surrounding the photon direction
be less than 5\GeV.
This selection is found to be equivalent to the experimental requirements for
isolated photons described in Section \ref{sec:photon_reco}.
These events are then processed through the full CMS detector simulation
chain using the
\GEANTfour{} package.
In order to model the effect of the underlying PbPb events, the
\PYTHIA\ photon events are embedded
into background events generated using
\HYDJET{} (v 1.8) \cite{Lokhtin:2005px}. This version of \HYDJET\ is tuned to reproduce
event properties such as charged hadron multiplicity, \pt\ spectra,
and elliptic flow measured
as a function of centrality in PbPb collisions.

\subsection {Photon reconstruction and identification}
\label{sec:photon_reco}

Photon candidates are reconstructed from clusters of energy deposited in the ECAL,
following the method detailed in Ref.~\cite{HIPhoton}. The selected photon candidates are restricted
to be in the barrel region of the ECAL by requiring a
pseudorapidity limit of $|\eta^\gamma|<1.44$ and are also required to have a transverse momentum
of $\ptg > 60\GeVc$.
In addition, photon candidates are dropped if they overlap with any
electron tracks. Electrons are identified by matching tracks with reconstructed ECAL
clusters and putting a cut on the ratio of calorimeter energy over track momentum. The separation of the photon and electron is required to be within a search window of $|\eta^\gamma-\eta^{\rm
  Track}|<0.02$ and $|\phi^\gamma-\phi^\text{Track}|<0.15$.
Anomalous signals caused by the interaction of heavily-ionising particles directly with the silicon avalanche photodiodes used for the ECAL barrel readout are  removed, again using the prescription of Ref.~\cite{HIPhoton}.
The reconstructed photon energy is corrected to account for the
material in front of the ECAL and for electromagnetic shower containment.
An additional correction is applied to the clustered energy in order to
remove the effects from the PbPb underlying event (UE).
The size of the combined correction
is obtained from the isolated photon \PYTHIAHYDJET{} sample
and varies from $2$--$10\%$, depending on centrality and photon $\ptg$.
The effect of the corrections on the energy scale is validated by an analysis of the reconstructed \cPZ{}
boson mass observed in $\cPZ \rightarrow \Pem\Pep$ decays in PbPb data as a
function of centrality.

Since the dominant source of neutral mesons is jet fragmentation with
associated hadrons, a first rejection of neutral mesons mimicking a high-\pt photon
in the ECAL is done using the ratio of hadronic to electromagnetic
energy, $H/E$. The $H/E$ ratio is calculated using the energy depositions in the
HCAL and the ECAL
inside a cone of $\Delta R = 0.15$ around
the photon candidate direction~\cite{CMSppPhoton}. Photon candidates with $H/E <
0.1$ are selected for this analysis. A correction for the contribution from the
remaining short-lived neutral mesons is applied later.

To determine if a photon candidate is isolated, the detector activity in a cone
of radius $\Delta R = 0.4$ with respect to the centroid of
the cluster is used. The UE-subtracted photon isolation variable
${\rm SumIso^{\text{UE--sub}}}$, which is the sum of transverse energy
measured in three sub-detectors (ECAL, HCAL, Tracker) minus
the expected contribution from the UE to each sub-detector,
 as described in~\cite{HIPhoton}, is used to further reject
photon candidates originating from jets.
The mean of ${\rm SumIso^{\text{UE--sub}}}$ for fragmentation and
decay photons is $\approx 20\:\mathrm{GeV}$, while the distributions of ${\rm SumIso^{\text{UE--sub}}}$ for isolated
photons are Gaussians centred around $0$ and having widths
varying from $3.5\:\mathrm{GeV}$ for peripheral collisions to
$8.5\:\mathrm{GeV}$ for central collisions.
Candidates with ${\text{SumIso}^{\text{UE--sub}}}$
smaller than 1\GeV are selected for further study. A tightened
isolation criterion for data (as compared to the 5~GeV applied for the MC) is used in order to minimise the
impact of random \PbPb\ UE
fluctuations. A downward fluctuation in the UE contribution to ${\text{SumIso}^{\text{UE--sub}}}$ can inadvertently
allow a non-isolated photon candidate to pass the isolation cut. From
the \PYTHIAHYDJET{} sample, the fraction of photons surviving
this tightened selection
is estimated to be $70$--$85\%$,
depending on centrality and photon \pt{}, and is found not to be
dependent on the angular or momentum correlation with the associated
jet. The relative efficiency of ${\text{SumIso}^{\text{UE--sub}}} <
1\:\mathrm{GeV}$ compared to ${\text{SumIso}^{\text{UE--sub}}} <
5\:\mathrm{GeV}$ ranges from $82\%$ ($0$--$10\%$ centrality) to $90\%$
($50$--$100\%$ centrality). At the same time, the photon purity,
central to peripheral, is $74$--$83\%$ for the
${\text{SumIso}^{\text{UE--sub}}} < 1\:\mathrm{GeV}$ cut, compared to
$52$--$62\%$ for the ${\text{SumIso}^{\text{UE--sub}}} <
5\:\mathrm{GeV}$ cut.

Photon purities in each centrality interval are estimated using a two-component fit of the shape of the electromagnetic
shower in the ECAL, \sigeta, defined as a modified second moment of the
electromagnetic energy cluster distribution around its mean $\eta$ position:
\begin{equation}
\label{sieieFormula}
\begin{split}
\sigma_{\eta \eta}^2 = \frac{\sum_i w_i(\eta_i-\bar{\eta})^2}{\sum_i
  w_i}, \\
   w_i = \mathrm{max}\left(0, 4.7 + \ln \frac{E_i}{E}\right),
  \end{split}
\end{equation}
where $E_i$ and $\eta_i$ are the energy and position of the $i$-th ECAL crystal in a group of $5\times 5$ crystals
centred on the one with the highest energy, $E$ is the total energy of the
crystals in the calculation, and $\bar{\eta}$ is the average
$\eta$ weighted by $w_i$ in the same group~\cite{CMSppPhoton}. The discrimination
is based only on the pseudorapidity (i.e.\ longitudinal) distribution
of the shower, which is aligned with
the magnetic field direction. As a result, showers with a wider distribution in the transverse plane, which can originate from
photons converted to $\Pep\Pem$ pairs in the detector material, are not eliminated.
The shape of the \sigeta\ distribution for the signal
is obtained from \photonjet\ \PYTHIAHYDJET{} samples
for each $p^\gamma_\mathrm{T}$ and centrality bin. The shape of the background distribution
is extracted from data using a background-enriched set of
photon candidates with $10<\text{SumIso}^{\text{UE--sub}}<20\GeV$. The
estimated photon purity (one minus the nonphoton contamination) is $74$--$83\%$
for photon candidates, which are required to have $\sigma_{\eta\eta}<0.01$.

\subsection{Jet reconstruction}
\label{sec:jet_reco}

Jets are reconstructed by clustering particles measured with a
particle-flow (PF) algorithm~\cite{CMS-PAS-PFT-10-002}, using the \ak\ sequential
recombination algorithm with a distance parameter of $R = 0.3$~\cite{Cacciari:2008gp}.
The jets used in the analysis are required to have $\text{\ptj} >
30\GeVc$ and $\lvert\eta^\text{Jet}\rvert<1.6$ to ensure high
reconstruction efficiency.
Jets within $R < 0.3$ around a photon are removed in order not to
correlate the photon with itself.
Details of the jet reconstruction procedure and its performance can be found
in~\cite{CMS:2012ni}.
The small value of $R$, compared to a more typical $R = 0.5$--$0.7$ used
to analyse \pp{} events, helps to minimise sensitivity to the UE contribution, and
especially its fluctuations.
The energy from the UE is subtracted using the same method
as employed in \cite{Chatrchyan:2011sx, CMS:2012ni} and originally described in~\cite{Kodolova:2007hd}.
The jet energy resolution can be quantified using the Gaussian standard deviation $\sigma$ of
$\pt^{\text{Reco}}/\pt^{\text{Gen}}$, where $\pt^{\text{Reco}}$ is the
UE-subtracted, detector-level jet energy, and $\pt^{\text{Gen}}$
is the generator-level jet energy without any contributions from a PbPb UE. The magnitude of this
resolution is determined using \PYTHIAHYDJET{} simulation
propagated through the detector using $\GEANTfour$. Compared to direct
embedding into \PbPb{} events, this method avoids uncertainties
associated with the detector versus MC geometry alignment, which is
especially difficult to achieve accurately with finely segmented pixel
trackers. The UE produced by \HYDJET{} with $\GEANTfour$ has
been checked against the data by observing the energy collected inside randomly oriented
cones with the same radius as the distance parameter in the jet
algorithm. Data and MC are found to be well matched.
The dependence of the jet resolution, $\sigma$, defined as the standard
deviation of the reconstructed over the event generator \ptj{}, can be parametrised using the expression
\begin{equation}
\sigma\left(\frac{\pt^{\text{Reco}}}{\pt^{\text{Gen}}}\right)
= C \oplus \frac{S}{\sqrt{\pt^{\text{Gen}}}} \oplus \frac{N}{\pt^{\text{Gen}}},
\label{eq:jet_energy_scale_parametrisation}
\end{equation}
where $\oplus$ indicates a sum in quadrature, and the quantities $C$, $S$, and $N$ are
fitted parameters (Table \ref{tab:jet_energy_scale_parametrisation}). The first
two terms of the parametrisation are determined from \PYTHIA{} simulation, and
the third term, which represents background fluctuations (not
corrected for the flow direction), is determined from \PYTHIAHYDJET{}
simulation.

Because the effects of the UE for jets found in \PbPb\ events are subtracted,
corrections to the mean reconstructed jet energy are derived from \pp{} data and $\PYTHIA$-only
simulation (i.e.\ without $\HYDJET$)~\cite{Chatrchyan:2011ds}.
Studies of the performance of jet reconstruction in \PYTHIAHYDJET{} events
show that no additional centrality-dependent energy correction is needed. %an additional centrality-dependent energy correction is not needed.

The jet reconstruction efficiency is defined as the fraction of
simulated \PYTHIA{} jets which are correctly reconstructed when embedded into a \HYDJET{} event.
The efficiency is found to be greater than $90\%$ for jets within the selected $\pt$ and $\eta$ range for all
centralities.
For the analysis of the \pp\ sample, the same PbPb jet reconstruction algorithm is used. The performance of the jet reconstruction in peripheral PbPb events is found to approach that for the \pp\ simulation.

\begin{table*}[htbp]
  \topcaption{Parameters of the functional form for the jet energy resolution
    $\sigma\left(\pt^{\text{Reco}}/\pt^{\text{Gen}}\right)$ given in
    Eq.~\eqref{eq:jet_energy_scale_parametrisation}, obtained from
    \GEANTfour{} simulation of \PYTHIA{} \pp{} jets and from
    \PYTHIA{} jets embedded in \HYDJET{} events for various \PbPb{}
    centralities (indicated by the \% ranges in parentheses). The units of $S$ are $\sqrt{\GeVc}$ and the units of $N$ are $\GeVc$.}
  \begin{center}
    \begin{tabular}{c|c|c|c|c|c|c}
      $C$ & $S$ & $N$ (\pp{}) & $N$ (50--100\%)& $N$ (30--50\%)& $N$ (10--30\%)& $N$ (0--10\%) \\
      \hline
      $0.0246$ & $1.213$ & $0.001$ & $0.001$ & $3.88$ & $5.10$ & $5.23$ \\
    \end{tabular}
  \end{center}
  \label{tab:jet_energy_scale_parametrisation}
\end{table*}

\subsection{Analysis procedure}
\label{sec:analysis}

To construct \photonjet\ pairs, the
highest $\ptg$ isolated photon candidate
in each selected event is associated with every
jet in the same event.
The \photonjet\ pairs constructed in this way contain background contributions
that need to be subtracted before using them to study energy loss effects on the jet produced in the same scattering as the photon.
The dominant background contributions are photons from meson decays which pass the isolation requirement and the combinatoric background where the leading photon is paired with a jet not originating from the same hard scattering.
The combinatoric background includes misidentified jets
which arise from fluctuations of the underlying event
as well as real jets from multiple hard interactions in the collision.

The background contributions from decay photon and fake jets are estimated separately with methods that are
data-driven and are subtracted from the \photonjet\ pair sample.

The estimation of the yield and the kinematic characteristics of decay photons contained in the isolated-photon sample
is based on the shower shape distributions for the analysed ECAL clusters.
The ECAL clusters originating from high-\pt\ meson decays correspond to two photons that
are reconstructed as a single wide cluster.
Events with a large shower width ($0.011 < \sigeta < 0.017$, see Eq.~\eqref{sieieFormula} are used to determine the contributions of the decay photon background to the $\dphijg$ and $\xjg$ observables.
The background shape obtained
from this procedure is scaled
according to the background-photon fraction, which is estimated from a fit of
the shower shape distribution. The estimated background contribution
fraction (which is equal to $1 - \text{purity}$) is then
subtracted from the yield for the
signal events, which have a small shower width ($\sigeta < 0.01$).

The background contribution due to \photonjet\ pairs arising from fake jets
or multiple hard scatterings is also subtracted. It is estimated by correlating each isolated
highest-$\pt$ photon from the
triggered \photonjet\ sample to jets found in a different event selected randomly from a
set of minimum bias \PbPb\ data.
The random event used in the pairing is chosen to have the same centrality as the
\photonjet\ candidate event.
The fake jet background estimated in this way has a flat distribution in \dphijg{}. The effect of this background is biggest in the most central events where, on average, approximately 20\% of the jets paired with each photon candidate are estimated to be fake jets.
The estimated distributions of $\dphijg$ and $\xjg$ for photons paired with fake jets, found using this random pairing
of events,
are subtracted from the distributions
coming from the same-event \photonjet\ sample to obtain the final results.

To determine the sensitivity to a potentially modified jet
fragmentation, which may cause the reconstructed jet energy scale to
deviate from the \PYTHIA{} derived calibration, the MC studies were
repeated but now using \PYQUEN{} \cite{Lokhtin:2005px} jets embedded
into \HYDJET{}. \PYQUEN{} simulates parton energy loss by radiative
and collisional mechanisms, where a portion of the original parton
energy is redistributed into gluons that are found largely outside the
cone of the surviving jet. The \PYQUEN{}+\HYDJET{} events were run
through the full detector simulation and then reconstructed with the
standard analysis. The jet modification in \PYQUEN{} produces a
\photonjet\ momentum imbalance comparable to that observed in our
measurement (although in detail, with different \xjg{} distribution
and \npart{} dependence). The extracted momentum imbalance was found
to reproduce the generator level imbalance well within the statistical
uncertainties. We also note that a similar insensitivity to
differences among QCD fragmentation was found previously by studying
the jet energy scale from separate \PYTHIA{} gluon and light quark
jets, which differ significantly in their fragmentation patterns
\cite{MattPFlow}. The standard analysis using PF jets was cross 
checked using jets reconstructed with only information from the 
calorimeters. This alternative analysis has different corrections 
for jet energy scale and resolution and a different sensitivity to 
low momentum tracks. The two analyses give comparable results for the 
photon+jet observables.

\section{Results}
\label{sec:results}

\subsection{Photon+jet azimuthal correlations}

\begin{figure*}[htbp]
\begin{center}
\includegraphics[width=0.99\textwidth]{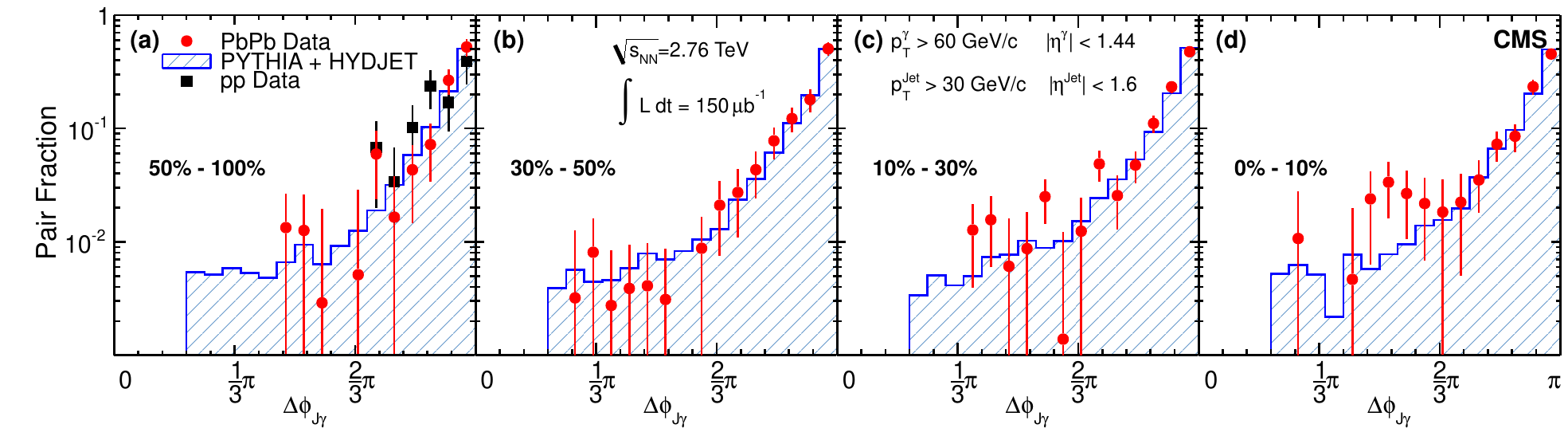}
\caption{\label{fig:gJdeltaPhi_qcdPhoRef_pp2760-xJ30G60} Azimuthal correlation $\dphijg$ between the
  photon and associated jet after background subtraction.
  The area of each distribution is normalised to unity. All panels show
\PbPb\ data (filled circles) compared to \pp\ data at
  2.76\TeV (filled squares), and to the \PYTHIAHYDJET{} MC simulation
  (shaded histogram) in bins of increasing centrality left to right. The error bars
  on the points represent the statistical uncertainty.
}
\end{center}
\end{figure*}

\begin{figure}[htbp]
\begin{center}
\includegraphics[width=0.48\textwidth]{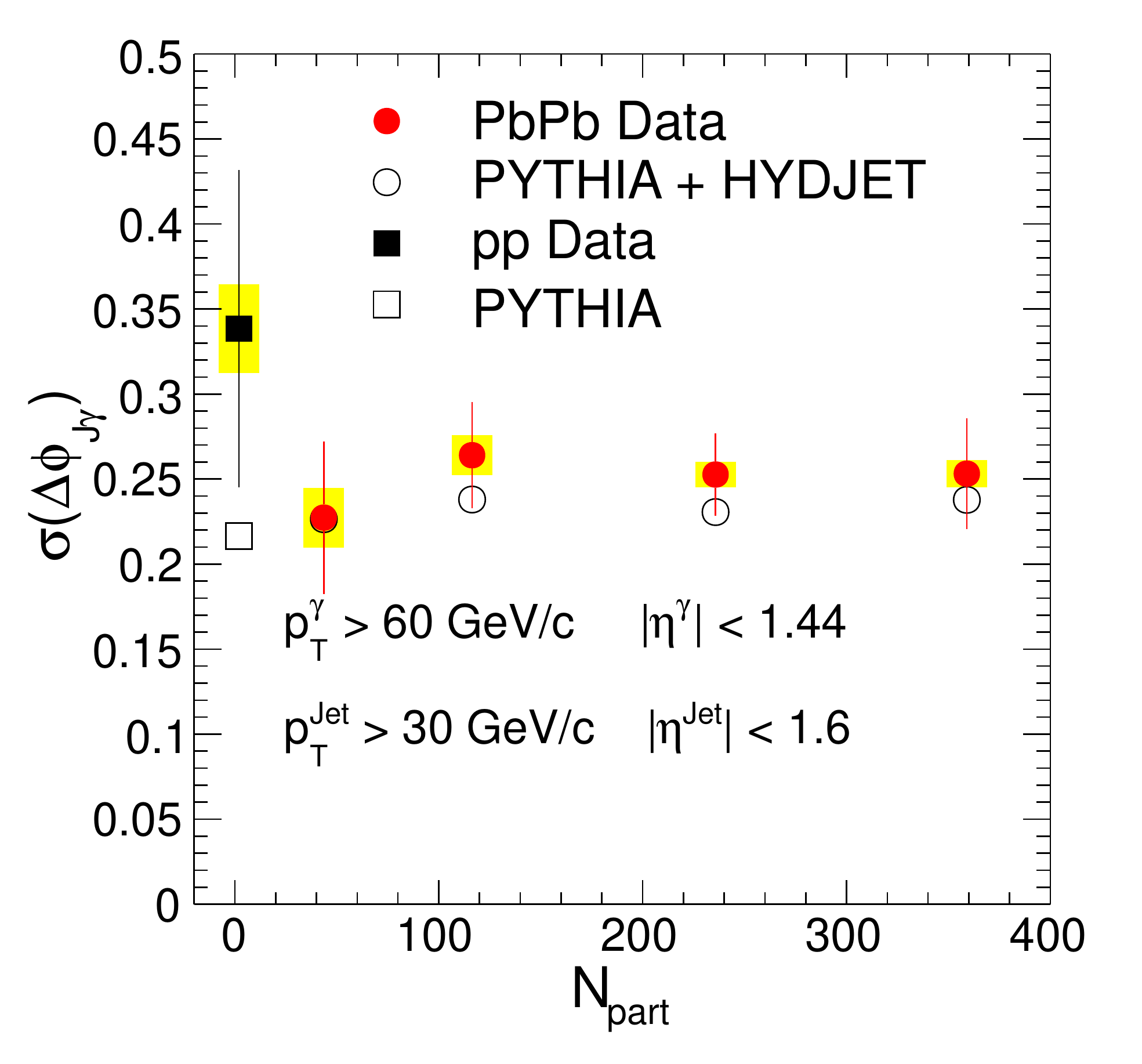}
\caption{\label{fig:resultsummarya}
Fitted \dphijg\ width ($\sigma$ in Eq.~\eqref{eq:sig}) between the photon and associated jet after background
subtraction as a function of $\npart$. The fit range was restricted to
$\dphijg > \frac{2}{3}\pi$.
The yellow boxes indicate point-to-point systematic uncertainties and the error bars
denote the statistical uncertainty.
}
\end{center}
\end{figure}

Possible medium effects on the back-to-back alignment of
the photon and recoiling jet can be studied using the
distribution of the number of \photonjet\ pairs, $N_{J\gamma}$, as a function of the relative azimuthal
angle, $\dphijg$, normalised the total number of pairs, $(N_{J\gamma})^{-1} \rd N_{J\gamma}/\rd\dphijg$.
Figure~\ref{fig:gJdeltaPhi_qcdPhoRef_pp2760-xJ30G60} shows distributions of \dphijg\ for PbPb data in four centrality bins, ranging
from peripheral events
(50--100\%, Fig.~\ref{fig:gJdeltaPhi_qcdPhoRef_pp2760-xJ30G60}a) to the most central events (0--10\%, Fig.~\ref{fig:gJdeltaPhi_qcdPhoRef_pp2760-xJ30G60}d).
The \PbPb\ data are compared to \PYTHIAHYDJET{} simulation and \pp\ data.
For both \PbPb\ data and MC distributions, the jet is found to be well aligned opposite to the photon direction,
with a clear peak at $\dphijg = \pi$. The shape of the $\dphijg$ correlation peak is similar
in \PbPb\ data and MC.
The apparent excess in the tail of the 0--10\% data was investigated and deemed statistically not significant compared to the subtracted background.
To study the centrality evolution of the
shape, the distributions are fitted
to a normalised exponential function:
\begin{equation}
\label{eq:sig}
  \frac{1}{N_{J\gamma}}
  \frac{\rd N_{J\gamma}}{\rd\dphijg} =
  \frac{\re^{(\Delta\phi - \pi)/\sigma}}{(1 -
    \re^{-\pi/\sigma})\,\sigma}.
\end{equation}
The fit is restricted to the exponentially falling region $\dphi > 2\pi/3$.
The results of this fit for \PbPb\ data are shown in
Fig.~\ref{fig:resultsummarya}, where
the width of the azimuthal correlation ($\sigma$ in Eq.~\eqref{eq:sig}, denoted \sjg\ in
Fig.~\ref{fig:resultsummarya}) is plotted as a function of centrality and
compared to \pp\ and \PYTHIAHYDJET{} fit results.
The resulting  \sjg\ values in \PbPb\ do not show a significant centrality dependence within the
present statistical and systematic uncertainties. For central \PbPb\
collisions, \sjg\ is similar to the \PYTHIA\ reference based on the
Z2 tune, and comparison with other \PYTHIA\ tunes shows a theoretical uncertainty
that is larger than the difference between the data and MC.
Comparing the \PYTHIA{} tune Z2 with tune D6T~\cite{Field:2008zz,Bartalini:2011xj} shows an 8\% difference in
\sjg{}, which is expected because these two tunes differ in their parton shower
ordering resulting in a different $\Delta\phi$ correlation.
The large statistical uncertainty in the \sjg{} extracted from the \pp\  data at 2.76\TeV does not allow
a discrimination between these two \PYTHIA tunes.
Both the Z2 and D6T tunes matched the shape of the azimuthal dijet correlation measured in \pp{}
collisions at 7\TeV~\cite{Khachatryan:2011zj} at about the 10\% level in the region $\dphi > 2\pi/3$.
The result that \sjg\ is not found to be significantly modified by the medium is consistent with the
earlier observation of an unmodified $\Delta\phi$ correlation in dijet events \cite{Chatrchyan:2011sx}.

\subsection{Photon+jet momentum imbalance}

\begin{figure*}[htbp]
\begin{center}
\includegraphics[width=0.95\textwidth]{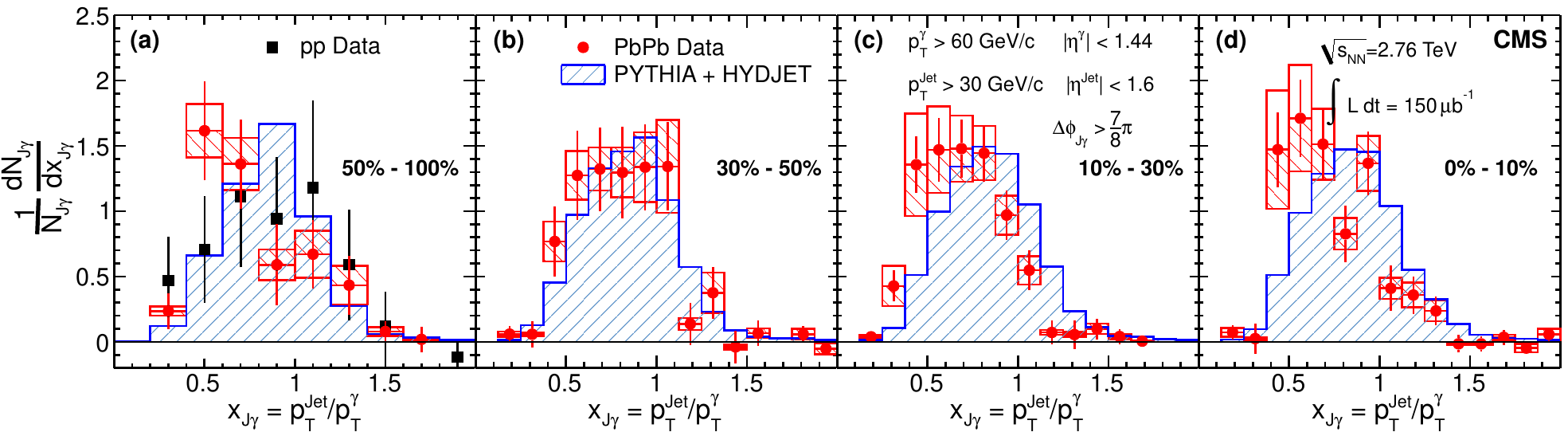}
\caption{\label{fig:InclPtRatio_qcdPhoRef_pp2760-xJ30G60} Ratio of \pt{} between the
  photon ($\pt^{\gamma} > 60$\GeVc) and jet ($\pt^{\text{Jet}} > 30$\GeVc, $\dphijg > \frac{7}{8}\pi$)
  after subtracting background. The area of each distribution is normalised to unity. All panels show
\PbPb\ data (filled circles) compared to \pp\ data at
  2.76\TeV (filled squares), and to the \PYTHIAHYDJET{} MC simulation
  (shaded histogram) in bins of increasing centrality left to right. The error bars
  on the points represent the statistical uncertainty.
  See text for an explanation of the open and shaded red systematic
  uncertainty boxes.
}
\end{center}
\end{figure*}

The asymmetry ratio
$\xjg= \ptj/\ptg$ is used to quantify the \photonjet\ momentum imbalance. In addition to the jet and photon selections used in
the \dphijg\ study, we further impose a strict $\dphijg >
\frac{7}{8}\pi$ cut to suppress contributions from background jets.
Note that \photonjet\ pairs for which the associated jet falls below the $30\GeVc$
threshold are not included in the $\xjg$ calculation. This limits the bulk of the $\xjg$ distribution to $\xjg \gtrsim 0.5$.
Figure~\ref{fig:InclPtRatio_qcdPhoRef_pp2760-xJ30G60} shows the centrality dependence of \xjg\ for
\PbPb collisions as well as that for \PYTHIAHYDJET{}
simulation where \PYTHIA{} contains inclusive isolated photon
processes. The \avexjg{} obtained from \PYTHIA{} tunes Z2 and D6T
agree to better than 1\%.
Overlaid in the peripheral bin is the \avexjg\ for 2.76\TeV \pp\ data,
showing consistency to the MC reference. However the poor statistics
of the \pp\ data and the 50--100\% \PbPb{} centrality bin do not offer a strong constraint on a specific MC reference.
However, further studies using the 7\TeV high statistics \pp\ data showed a
good agreement in \avexjg{} between data and \PYTHIA{}, justifying the
use of \PYTHIAHYDJET{} as an un-modified reference.
The dominant source of systematic uncertainty in \avexjg{} is
the relative \photonjet{} energy scale. Its impact on the probability
density of $\xjg$ is approximately 10\%
for the intermediate region of $0.6 < \xjg < 1.2$.
The normalisation to unity causes a point-to-point anticorrelation in the
systematic uncertainties, where the upward movement of the probability
density at small $\xjg$ has to be offset by the corresponding downward movement
at large $\xjg$. This is represented by the
separate open and shaded red systematic uncertainty boxes in Fig.~\ref{fig:InclPtRatio_qcdPhoRef_pp2760-xJ30G60}.
For a given change in the energy scale, all points would move together in the direction
of either the open or shaded red box.
The $\npart$ dependence of the mean value $\avexjg$ is shown in Fig.~\ref{fig:resultsummarybc}(a).

While the \photonjet\ momentum ratio in the \PYTHIAHYDJET{}
simulation shows almost no change in the peak location and only a modest broadening, even in the most central \PbPb{} events, the \PbPb{} collision data exhibit a
change in shape, shifting the distribution towards lower \xjg\ as a function of centrality. It is important to
note that, as discussed above, the limitation of $\xjg \gtrsim 0.5$ limits the degree to
which this distribution can shift.

\subsection{Jet energy loss}

\begin{figure}[htbp]
\begin{center}
\includegraphics[width=0.48\textwidth]{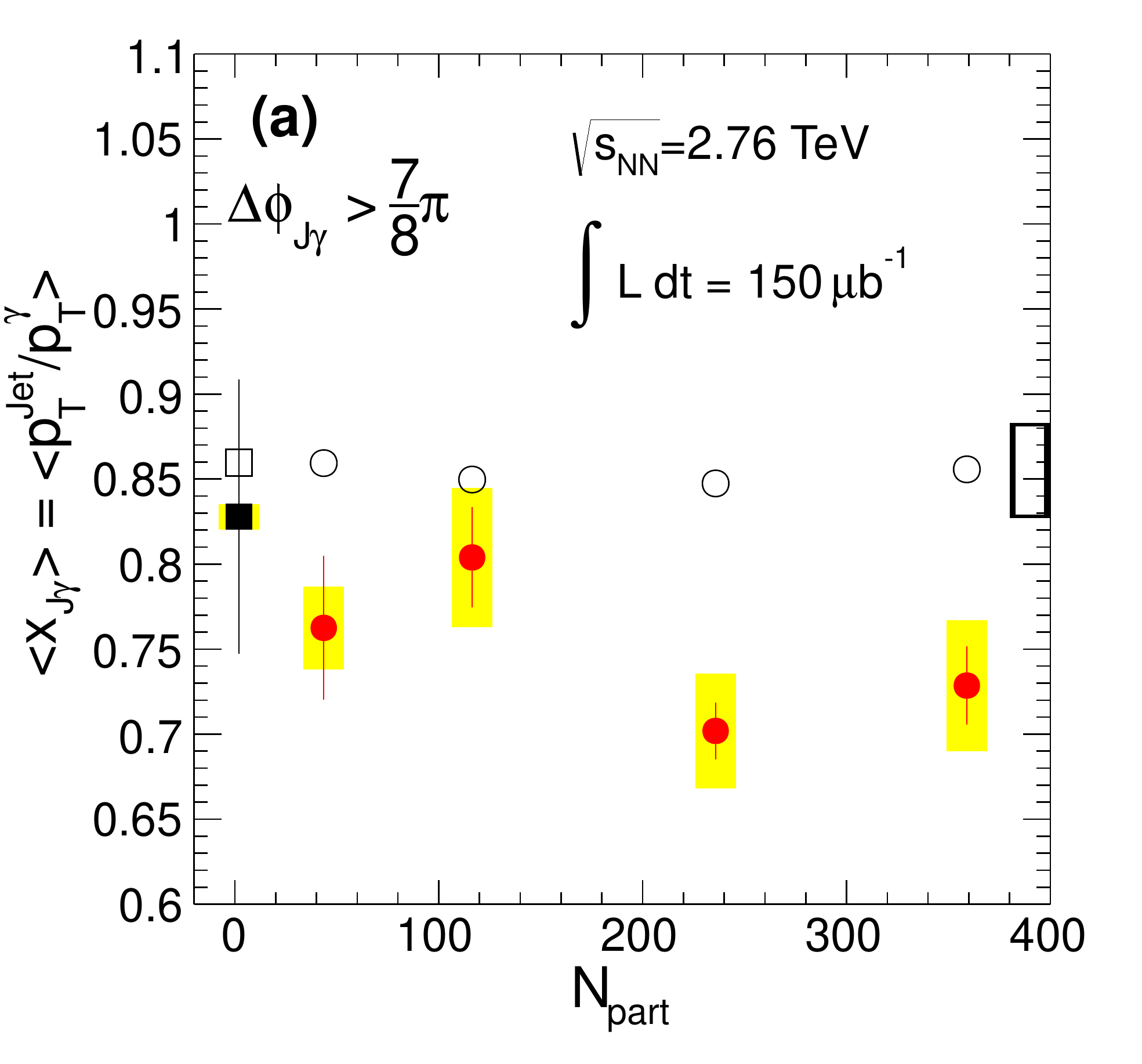}
\includegraphics[width=0.48\textwidth]{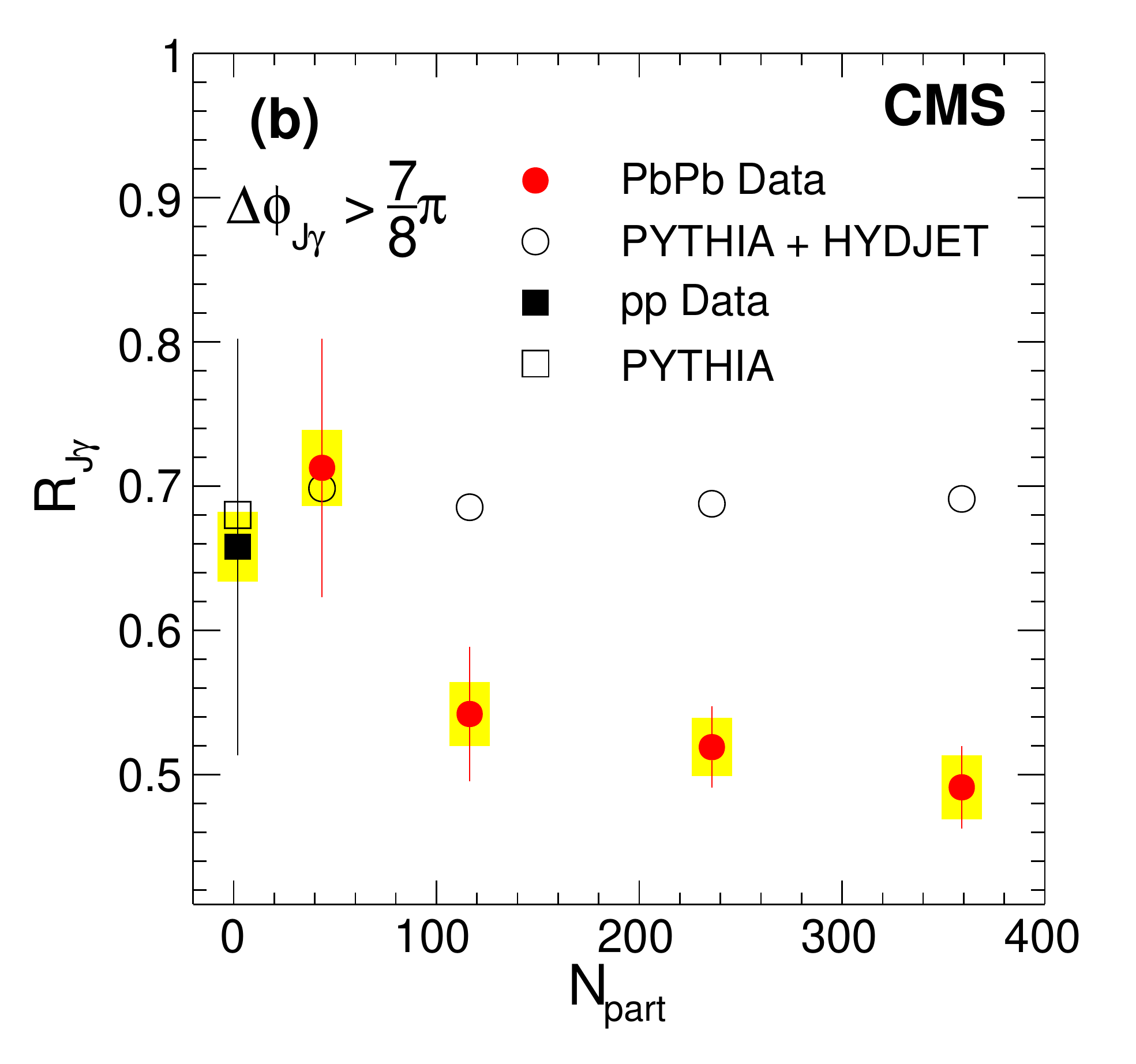}
\caption{\label{fig:resultsummarybc}
(a) Average ratio of jet transverse momentum to photon transverse
momentum, $\avexjg$, as a function of $\npart$. The empty box at the far right indicates
the correlated systematic uncertainty.
(b) Average fraction of isolated photons with an associated jet above
30\GeVc, $\rjg$, as a function of $\npart$.
In both panels, the yellow boxes indicate point-to-point systematic uncertainties and the error bars
denote the statistical uncertainty.
}
\end{center}
\end{figure}

To study the quantitative centrality evolution of the energy loss,
the average ratio of the jet and photon transverse momenta, \avexjg{},
is shown in Fig.~\ref{fig:resultsummarybc}(a).
While the \photonjet\ mean momentum ratio in the \PYTHIAHYDJET{} simulation exhibits a
roughly centrality-independent value of
$\avexjg = 0.847\pm 0.004\text{(stat.)}$ -- $0.859\pm
0.005\text{(stat.)}$, the ratio is $\avexjg = 0.73\pm
0.02\text{(stat.)}\pm 0.04\text{(syst.)}$ in the most
central \PbPb\ data, indicating that the presence of the medium results in more
unbalanced \photonjet\ pairs.

It is important to keep in mind that the average energy loss of the selected \photonjet\ pairs does not constitute the full picture.
There are genuine \photonjet\ events which do not contribute to the $\langle\xjg\rangle$ distribution
because the associated  jet falls below the $\ptj > 30\GeVc$ threshold.
To quantify this effect, Fig.~\ref{fig:resultsummarybc}(b) shows $\rjg$, the fraction of isolated photons that have an associated jet passing the analysis selection.
The value of $\rjg$ is found to
decrease, from $\rjg = 0.685\pm 0.008\text{(stat.)}$--$0.698\pm 0.006\text{(stat.)}$ for the \PYTHIAHYDJET{} reference,
as well as \pp{} and peripheral \PbPb{} data, to the significantly
lower $\rjg = 0.49\pm 0.03\text{(stat.)}\pm 0.02\text{(syst.)}$--$0.54\pm 0.05\text{(stat.)}\pm 0.02\text{(syst.)}$ for
the three PbPb bins
above 50\% centrality.

An analysis with a lower \pt{} cutoff on the associated jet energy would result in values of \rjg{} closer to unity. This would shift the cutoff at low \xjg{} in Fig.~3 closer to zero. It is likely, although not certain, that these additional events would result in a larger deviation in \xjg{} between the \PbPb{} data and the reference shown in Fig. 4(a).

\subsection{Systematic uncertainties}
\label{sec_syserr}

Photon purity, reconstruction efficiency, and isolation, as well as the contamination
from $e^\pm$ and fake jets contribute to the systematic uncertainties
of the \photonjet\ azimuthal correlation and the observables related to momentum asymmetry,
$\langle\xjg\rangle$ and \rjg{}.
Additionally, the momentum asymmetry observables are also influenced by
the relative photon and jet energy calibrations. For the measurement of $\sigma(\dphi)$, the
uncertainty due to the photon angular resolution is negligible, less than $10^{-5}$.

The uncertainty in the relative \photonjet\ energy scale consists of four main contributions.
The first one comes from the $2\%$
relative uncertainty of the jet energy scale in the barrel for
$30< \ptj < 200\GeVc$, when compared with the ECAL energy scale
\cite{Chatrchyan:2011ds}. The second contribution is the residual
data-to-MC energy scale difference in \pp{} collisions, which
is not corrected for in this analysis, for which we
quote the $2\%$ maximum relative uncertainty which applies in the range
$\lvert\eta^\text{Jet}\rvert < 1.6$.
Thirdly, the additional uncertainty for the jet energy scale in the
presence of the UE is determined to be $3\%$ for the $30$ to $100\%$ and $4\%$ for the
$0$ to $30\%$ centrality range, using the embedding of
\PYTHIA{} isolated \photonjet\ pairs into
\HYDJET{}. The fourth contribution is the effect of heavy ion background on the
ECAL energy scale, which is determined from $\cPZ \rightarrow \Pem\Pep$ mass
reconstruction, after applying the \PbPb{} ECAL correction. This results in a relative uncertainty of 1.5\%,
comparable to the \pp{} uncertainty (obtained via $\pi^0$ and $\eta
\rightarrow \cPgg\cPgg$).

The absolute photon energy scale uncertainty, estimated to be 1.5\% using \cPZ{} decays as described above,
will also affect the threshold of our
photon kinematic selection.
Similarly, the lower transverse momentum cutoff for jets is sensitive to
their absolute energy scale.
For CMS, the energy of jets is calibrated by measuring
the relative \photonjet{} energy scale in \pp\ collisions, %(assuming no modification of the jet energy),
and therefore the uncertainty in jet energies is the quadrature
sum of the uncertainties in the relative jet-to-photon energy scale and the absolute photon energy scale.

The uncertainty of the photon purity measurement using the  \sigeta\ template
fitting is estimated by (a) varying the selection of sideband
regions that is used to obtain the background template and (b) shifting
the template to measure the signal template uncertainty.
These result in an estimated uncertainty on the photon purity of 12\% and 2\%, respectively.
Systematic effects due to photon reconstruction efficiency are estimated by correcting the data using the efficiency derived from the MC simulation,
and comparing the result with the uncorrected distribution.  The contribution of non-isolated photons
(mostly from jet fragmentation) that are incorrectly determined to be
isolated in the detector due to UE energy fluctuations
or detector resolution effects is
estimated using \PYTHIAHYDJET{} simulation.  The difference of
\photonjet{} observables obtained from generator level isolated
photons and detector level isolated photons is taken to be the systematic uncertainty
resulting from the experimental criterion for an isolated photon.

The current analysis removes contamination from fake jets
purely by subtracting the background estimated from event mixing.
A cross-check of this subtraction
has been performed using a direct rejection of fake jets via a fake
jet discriminant. The discriminant sums the $\pt^2$ of the jet core
within $R < 0.1$ around the jet axis and determines the likelihood that the
reconstructed jet is not the result of a background fluctuation. Both
techniques for fake jet removal agree within $1\%$ for the observables studied.
The effect of inefficiencies in the jet finding is
estimated by repeating the analysis and weighting each jet with the
inverse of the jet
finding efficiency as a function of \ptj.

\begin{table*}[htbp]
  \topcaption{Relative systematic uncertainties for
    $\sigma(\dphijg)$ for \pp\ data and each of the \PbPb\ centrality
    bins.}
  \begin{center}
    \begin{tabular}{l|c|c|c|c|c}
      Source & \pp & \PbPb{} 50--100\% & \PbPb{} 30--50\% & \PbPb{} 10--30\% & \PbPb{} 0--10\%\\
      \hline
      \cPgg{} \pt{} threshold & 3.0\% & 3.0\% & 3.0\% & 2.0\% &
      1.2\%\\
      Jet \pt{} threshold & 1.3\% & 1.3\% & 0.2\% & 0.5\% & 2.4\%\\
      \cPgg{} efficiency & 0.8\% & 0.8\% & 0.3\% & 0.3\% & 0.3\%\\
      Jet efficiency & 0.6\% & 0.6\% & 0.7\% & 0.4\% & 0.3\%\\
      Isolated \cPgg{} definition & 0.7\% & 0.7\% & 1.6\% & 2.0\% & 0.5\%\\
      \cPgg{} purity & 6.8\% & 6.8\% & 2.7\% & 0.5\% & 0.9\%\\
      $\Pem,\Pep$ contamination & 0.5\% & 0.5\% & 0.5\% & 0.5\% & 0.5\%\\
      Fake jet contamination & 0.3\% & 0.3\% & 0.1\% & 0.2\% & 1.2\%\\
      Jet $\phi$ resolution & 0.5\% & 0.5\% & 0.5\% & 0.5\% & 0.5\%\\
      $\sigma$ fitting & 0.3\% & 0.3\% & 0.1\% & 0.1\% & 0.1\%\\
      \hline
      Total & 7.7\% & 7.7\% & 4.5\% & 3.0\% & 3.2\%\\
    \end{tabular}
  \end{center}
  \label{tab:systematic_uncertainty_sigma_abs_delta_phi}
\end{table*}

\begin{table*}[htbp]
  \topcaption{Relative systematic uncertainties for $\langle
    \xjg\rangle$ for \pp\ data and each of the \PbPb\ centrality bins.
    The uncertainties
    due to the \pp{} \cPgg--jet relative energy scale and \cPgg{}
    purity are common to all of the measurements and are quoted as
    a correlated uncertainty.}
  \begin{center}
    \begin{tabular}{l|c|c|c|c|c}
      Source & \pp & \PbPb{} 50--100\% & \PbPb{} 30--50\% & \PbPb{} 10--30\% & \PbPb{} 0--10\%\\
      \hline
      \cPgg--jet rel. energy scale & 2.8\% & 4.1\% & 5.4\% & 5.0\% &
      4.9\%\\
      \cPgg{} \pt{} threshold & 0.6\% & 0.6\% & 0.6\% & 0.6\% &
      1.3\%\\
      Jet \pt{} threshold & 0.7\% & 0.7\% & 1.9\% & 1.9\% & 2.0\%\\
      \cPgg{} efficiency & $<0.1\%$ & $<0.1\%$ & $<0.1\%$ & 0.1\% &
      0.2\%\\
      Jet efficiency & 0.5\% & 0.5\% & 0.6\% & 0.6\% & 0.5\%\\
      Isolated \cPgg{} definition & 0.1\% & 0.1\% & 0.7\% & 0.4\% & 2.0\%\\
      \cPgg{} purity & 2.2\% & 2.2\% & 1.9\% & 2.4\% & 2.7\%\\
      $\Pem,\Pep$ contamination & 0.5\% & 0.5\% & 0.5\% & 0.5\% & 0.5\%\\
      Fake jet contamination & 0.1\% & 0.1\% & 0.1\% & 0.2\% & 0.1\%\\
      \hline
      Total & 3.7\% & 4.8\% & 6.2\% & 6.0\% & 6.4\%\\
      \hline
      Correlated (abs., rel.) & 3.6\% & 3.6\% & 3.6\% & 3.6\% &
      3.6\%\\
      Point-to-point & 0.9\% & 3.2\% & 5.1\% & 4.8\% & 5.3\%\\
    \end{tabular}
  \end{center}
  \label{tab:systematic_uncertainty_mean_x_j_gamma}
\end{table*}

\begin{table*}[htbp]
  \topcaption{Relative systematic uncertainties for the fraction of photons matched with jets, \rjg{},
for \pp\ data and each of the \PbPb\ centrality bins.}
  \begin{center}
    \begin{tabular}{l|c|c|c|c|c}
      Source & \pp & \PbPb{} 50--100\% & \PbPb{} 30--50\% & \PbPb{} 10--30\% & \PbPb{} 0--10\%\\
      \hline
      \cPgg{} \pt{} threshold & 2.0\% & 2.0\% & 1.9\% & 1.3\% & 2.1\%\\
      Jet \pt{} threshold & 1.4\% & 1.4\% & 2.3\% & 2.6\% & 2.7\%\\
      \cPgg{} efficiency & 0.2\% & 0.2\% & 0.2\% & 0.5\% & 0.5\%\\
      Jet efficiency & 1.5\% & 1.5\% & 1.7\% & 1.8\% & 2.1\%\\
      Isolated \cPgg{} definition & 0.2\% & 0.2\% & 0.6\% & 1.3\% & 0.8\%\\
      \cPgg{} purity & 2.3\% & 2.3\% & 1.9\% & 0.2\% & 0.9\%\\
      $\Pem,\Pep$ contamination & 0.5\% & 0.5\% & 0.5\% & 0.5\% & 0.5\%\\
      Fake jet contamination & 0.4\% & 0.4\% & 0.8\% & 1.0\% & 1.4\%\\
      \hline
      Total & 3.7\% & 3.7\% & 4.1\% & 3.9\% & 4.5\%\\
    \end{tabular}
  \end{center}
  \label{tab:systematic_uncertainty_r_x_gt_0}
\end{table*}

Tables~\ref{tab:systematic_uncertainty_sigma_abs_delta_phi},
\ref{tab:systematic_uncertainty_mean_x_j_gamma}, and
\ref{tab:systematic_uncertainty_r_x_gt_0} summarise the relative
systematic uncertainties for $\sigma(\dphi)$, $\avexjg$, and $\rjg$, respectively, for the \pp\
data and for each of the \PbPb\ centrality bins used in the analysis.
For $\avexjg$, the uncertainties are separated
into a correlated component that is common to all centrality bins and a component
that represents the point-to-point systematic uncertainty. The common correlated
uncertainty is obtained by combining the \pp{} jet energy scale
uncertainty with the photon purity uncertainty. This absolute uncertainty of 3.6\% was used as the
correlated uncertainty for all \PbPb\ centrality bins.
\section{Conclusions}
\label{sec:conclusion}
The first study of isolated-photon+jet correlations in
  $\PbPb$ collisions at $\sqrt{s_{NN}} = 2.76\TeV$ has been performed as a function of collision centrality using
a dataset corresponding to an integrated luminosity of 150\mubinv. Isolated
photons with $\ptg > 60$\GeVc were correlated with jets with $\ptj > 30$\GeVc to
determine the width of the angular
correlation function, \sjg, the jet/photon transverse momentum ratio, $\xjg = \ptj/\ptg$, and
the fraction of photons with an associated jet, \rjg.
The \PbPb{} data were compared to both \pp{} data and a \PYTHIAHYDJET{} MC reference which included the effect of the underlying \PbPb event
but no parton energy loss.
No angular broadening was observed beyond that seen in the pp data and MC reference at all centralities.
The average transverse momentum ratio for the most central events was found to be
$\avexjg_{0-10\%} = 0.73 \pm 0.02 \mbox{(stat.)} \pm 0.04 \mbox{(syst.)}$.
This is lower than the value of 0.86 seen in the pp data and predicted by \PYTHIAHYDJET{} at the same centrality.
In addition to the shift in
momentum balance, it was found that, in central \PbPb\ data, only
a fraction equal to $\rjg = 0.49 \pm 0.03 \text{ (stat.)} \pm 0.02 \text{ (syst.)}$ of photons are matched with an
associated jet at $\dphijg > \frac{7}{8} \pi$, compared to a value of 0.69 seen
in \PYTHIAHYDJET{} simulation. Due to the hot and dense medium created in central \PbPb\ collisions, the energy loss of the associated parton causes the corresponding reconstructed jet to fall
below the $\ptj > 30$\GeVc threshold for an additional 20\% of the selected photons.

\section*{Acknowledgments}
We congratulate our colleagues in the CERN accelerator departments for the excellent performance of the LHC machine. We thank the technical and administrative staff at CERN and other CMS institutes, and acknowledge support from: FMSR (Austria); FNRS and FWO (Belgium); CNPq, CAPES, FAPERJ, and FAPESP (Brazil); MES (Bulgaria); CERN; CAS, MoST, and NSFC (China); COLCIENCIAS (Colombia); MSES (Croatia); RPF (Cyprus); MoER, SF0690030s09 and ERDF (Estonia); Academy of Finland, MEC, and HIP (Finland); CEA and CNRS/IN2P3 (France); BMBF, DFG, and HGF (Germany); GSRT (Greece); OTKA and NKTH (Hungary); DAE and DST (India); IPM (Iran); SFI (Ireland); INFN (Italy); NRF and WCU (Korea); LAS (Lithuania); CINVESTAV, CONACYT, SEP, and UASLP-FAI (Mexico); MSI (New Zealand); PAEC (Pakistan); MSHE and NSC (Poland); FCT (Portugal); JINR (Armenia, Belarus, Georgia, Ukraine, Uzbekistan); MON, RosAtom, RAS and RFBR (Russia); MSTD (Serbia); MICINN and CPAN (Spain); Swiss Funding Agencies (Switzerland); NSC (Taipei); TUBITAK and TAEK (Turkey); STFC (United Kingdom); DOE and NSF (USA).

\bibliography{auto_generated}   % will be created by the tdr script.

\cleardoublepage \appendix\section{The CMS Collaboration \label{app:collab}}\begin{sloppypar}\hyphenpenalty=5000\widowpenalty=500\clubpenalty=5000\textbf{Yerevan Physics Institute,  Yerevan,  Armenia}\\*[0pt]
S.~Chatrchyan, V.~Khachatryan, A.M.~Sirunyan, A.~Tumasyan
\vskip\cmsinstskip
\textbf{Institut f\"{u}r Hochenergiephysik der OeAW,  Wien,  Austria}\\*[0pt]
W.~Adam, T.~Bergauer, M.~Dragicevic, J.~Er\"{o}, C.~Fabjan, M.~Friedl, R.~Fr\"{u}hwirth, V.M.~Ghete, J.~Hammer\cmsAuthorMark{1}, N.~H\"{o}rmann, J.~Hrubec, M.~Jeitler, W.~Kiesenhofer, V.~Kn\"{u}nz, M.~Krammer, D.~Liko, I.~Mikulec, M.~Pernicka$^{\textrm{\dag}}$, B.~Rahbaran, C.~Rohringer, H.~Rohringer, R.~Sch\"{o}fbeck, J.~Strauss, A.~Taurok, P.~Wagner, W.~Waltenberger, G.~Walzel, E.~Widl, C.-E.~Wulz
% ...
\vskip\cmsinstskip
\dag:~Deceased\\
1:~~Also at CERN, European Organization for Nuclear Research, Geneva, Switzerland\\
% ...

\end{sloppypar}
\end{document}